\documentclass[%
 reprint,
nofootinbib,
 amsmath,amssymb,
 aps,
pra,
prstper,
floatfix,
]{revtex4-2}

\usepackage{graphicx}
\usepackage{dcolumn}
\usepackage{bm}
\usepackage{hyperref}

\usepackage{physics}
\begin{document}

\preprint{APS/123-QED}

\title{Forward-backward correlations with the $\Sigma$ quantity in the wounded constituent framework at LHC energies}

\author{Iwona Sputowska}
\affiliation{H. Niewodniczański Institute of Nuclear Physics PAN, 31-342 Cracow, Poland}
\email{iwona.sputowska@ifj.edu.pl}

\date{\today}

\begin{abstract}

 $\Sigma$ is a new correlation measure, quite recently introduced to heavy-ion physics. This measure, defined in the independent source model as a strongly intensive quantity, is expected to be free of the effects of system volume and volume fluctuations. This paper discusses the forward-backward correlation quantified with the $\Sigma$ observable calculated in the framework of the wounded nucleon model (WNM) and wounded quark model (WQM). Findings show that the wounded constituent approach outperforms the commonly used heavy-ion Monte Carlo generators, such as HIJING, AMPT or EPOS, by accurately describing the experimental data on FB correlations with $\Sigma$ measured by ALICE in Xe--Xe reactions at $\sqrt{s_{\rm{NN}}}$=5.44 TeV and in Pb--Pb collisions at $\sqrt{s_{\rm{NN}}}$= 2.76 and 5.02 TeV. This paper demonstrates that $\Sigma$ can be a unique tool for determining the fragmentation function of a wounded constituent in a symmetric nucleus-nucleus collision. However, in the wounded constituent framework, it is no longer a strongly intensive quantity.

\end{abstract}

\maketitle


\section{\label{sec:level1}Introduction}

Over the last few years, there has been growing interest in the analysis of particle multiplicity correlations and fluctuations in high-energy nucleus-nucleus collisions with so-called $\emph{strongly intensive quantities}$.
These observables were first introduced in heavy-ion physics in Ref.~\cite{Gorenstein:2011vq} as a remedy to the spurious effect of volume (centrality) fluctuations that  contaminate  the measured physical variables such as the multiplicity correlation coefficient Refs.~\cite{Konchakovski:2008cf, Bzdak:2009xq}.  Strongly intensive quantities were defined in an $\emph{independent source model}$ framework  as observables that do not depend on system volume or its fluctuations  but rather carry direct information about a single (average) source producing particles. The term independent source model  refers to a class of superposition models that assume that particles are emitted independently from a collection of statistically identical sources. A fundamental example of the independent source model is the $\emph{wounded nucleon model}$ Ref.~\cite{Bialas:1976ed}, which assumes that a nucleus-nucleus collision can be  constructed  as a superposition of elementary nucleon-nucleon interactions. 

Two sets of strongly intensive quantities, $\Sigma$ and $\Delta$, were constructed as a combination of second moments and dedicated to studying correlations and fluctuations. Of the two, only the $\Sigma$ family contains a covariance term sensitive to particle correlations.

\begin{figure}[b]
\includegraphics[width=0.30\textwidth]{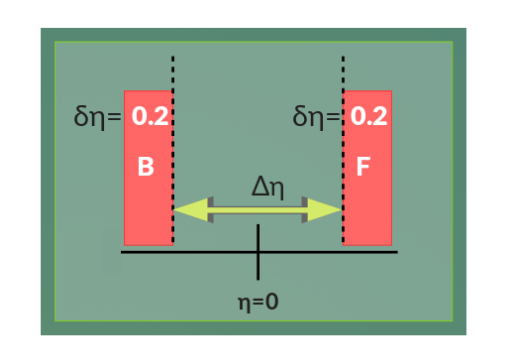}
\caption{\label{fig:FB_def}The figure illustrates the definition of forward F=$(\Delta\eta/2,~\Delta\eta/2+\delta\eta)$ and backward B=$(\Delta\eta/2 -\delta\eta,~-\Delta\eta/2)$ pseudorapidity intervals,  located symmetrically around $\eta=0$.}
\end{figure}

Recently, new results on forward-backward (FB) multiplicity correlations with strongly intensive quantity $\Sigma$ were measured in Pb--Pb collisions at $\sqrt{s_{\rm{NN}}}$ =2.76 and 5.02 TeV as well as Xe--Xe collisions at $\sqrt{s_{\rm{NN}}}$= 5.44 TeV in ALICE at LHC, Refs.~\cite{Sputowska:2019yvr, Sputowska:2022gai}. Results were obtained for pairs of forward F and backward B pseudorapidity ($\eta$) intervals of width $\delta\eta=$0.2, each located symmetrically around midrapidity ($\eta=0$) as illustrated in Fig.~\ref{fig:FB_def}. The $\Sigma$ observable values were studied for different centrality classes and as a function of the distance between the forward (F) and backward (B) pseudorapidity interval $\Delta\eta$.
In terms of this FB correlation analysis, $\Sigma$ quantity is defined by the Eq.~\ref{sigma_def}  as a combination of scaled variances  $\omega_{F(B)}$, first moments $\langle N_{F(B)}\rangle$, and covariance of multiplicity distributions:
\begin{equation}
\Sigma= \frac{ \omega_{B}\langle N_{F}\rangle+\omega_{F}\langle N_{B}\rangle -2\text{Cov}(N_{B}, N_{F})}{\langle N_{F}\rangle+\langle N_{B}\rangle} \label{sigma_def}.
\end{equation} 
ALICE Collaboration studies have confirmed that $\Sigma$ exhibits strongly intensive quantity properties in heavy-ion collisions. Specifically, it has been observed that it is independent of the centrality selection method in the experiment and free from the contribution from volume fluctuations within studied centrality classes, Ref.~\cite{Sputowska:2019yvr}.
It was found that $\Sigma$ displays a dependence on the energy and centrality of the collision system. The observed trend with  energy, namely the growth,  is similar for all studied nucleus-nucleus collision types. On the other hand,  the behavior of $\Sigma$ with centrality has the opposite character for collisions of larger systems such as Pb--Pb and Xe--Xe versus that observed in pp interactions, Refs.~\cite{Sputowska:2022CPOD, Sputowska:2022gai}. While theoretical predictions characterize the $\Sigma$ energy and centrality dependence in pp collisions reasonably well, the most commonly used heavy-ion Monte Carlo models fail to describe the data in Pb--Pb and Xe--Xe collisions. Models break down qualitatively (HIJING~\cite{HIJING}) and quantitatively (EPOS~\cite{EPOS}, AMPT~\cite{AMPT}) in their attempt to reproduce the $\Sigma$ measured in the experiment. Overall causal factors leading to the discrepancy between nucleus-nucleus and elementary pp collisions and between theoretical description and experimental results remain speculative and  are worth further investigation.

This paper examines the properties of the forward-backward strongly intensive quantity $\Sigma$ in the $\emph{wounded constituent model}$, Ref.~\cite{WCBialas:2012dc}. The constituent (also referred to as a source) can be either a nucleon or a quark participating in the collision. Four main conclusions can be drawn from this study. First, $\Sigma$, taken in terms of the wounded constituent model, gives a non-trivial result different from unity. Second, due to its intrinsic dependence on fluctuations of wounded nucleons, it is no longer a strongly intensive quantity. However, it preserves some strongly intensive properties as a result of its relation to partial correlations in symmetric collisions. Third, the wounded nucleon models and the wounded quark model explain qualitatively and quantitatively the observed experimental trends of $\Sigma$ with the centrality of the collision. Fourth, the $\Sigma$ variable appears to be responsive to the nature of the fragmentation function. The immediate inspiration for this paper came from the studies presented in Refs.~\cite{Bzdak:2009xq, Bzdak:2009dr,Barej:2017kcw}.

The remainder of this paper has been organized in the following way.
Section 2 defines the wounded nucleon and wounded quark models. The analytical formula for $\Sigma$ derived in their framework and the description of WNM and WQM Monte Carlo calculations are presented. In Section 3,  results are discussed and compared with the ALICE data. Based on the research findings, predictions are made regarding the characteristics of the wounded nucleon and wounded quark fragmentation function for LHC energies. 
The paper is closed with conclusions.

\section{ Forward-Backward strongly intensive quantities in wounded nucleon and wounded quark models}\label{s:WNMandWQM_def}

The superposition approach in which  wounded sources in the projectile and  target nuclei independently produce particles during the nucleus-nucleus collision was proposed in high-energy physics long ago, Ref.~\cite{Bialas:1976ed}.  In particular, the outcome of this theoretical framework depends on the postulated  nature of the sources (constituents). In this paper, two scenarios are considered: the wounded quark model and the wounded nucleon model, Refs.~\cite{Bialas:1976ed, Bialas:2004su, WQM,WCBialas:2012dc,Bialas_2008}.

The wounded nucleon model was found to be relatively successful in describing  soft particle production in relativistic heavy-ion collisions as a composition of independent contributions of nucleon-nucleon interactions, Refs.~\cite{Bialas:1976ed,WA97:2000kwf,Bzdak:2009dr, Bzdak:2009xq}. The main assumptions of the WNM, adopted in this paper in terms of forward-backward correlation analysis, can be characterized as listed below.
\begin{enumerate}
\item[1.] A wounded nucleon is a nucleon that has undergone at least one inelastic collision. Each wounded nucleon emits particles regardless of the number of collisions it went through.

\item[2.]The universal fragmentation function $F(\eta)$ describes the particle production for each wounded nucleon.

\item[3.]The fragmentation function F($\eta$) is not limited to the  hemisphere of the wounded nucleon. Any wounded source can emit particles into both forward and backward pseudorapidity regions.
\item[4.] A single particle density $N(\eta)\equiv\frac{dN(\eta)}{d\eta}$ in a nuclear collision can be expressed by the combination of inputs coming from the average number of  forward-going, $\langle w_{F}\rangle$,  and  backward-going, $\langle w_{B}\rangle$, wounded nucleons (see Ref.~\cite{Bialas:2004su}):
\begin{equation}
N(\eta)=\langle w_{F}\rangle F(\eta) +\langle w_{B}\rangle F(-\eta),
 \label{eq:particle_density}
\end{equation}

\item[5.] The number of particles produced in the collision fluctuates on an event-by-event basis according to a given multiplicity distribution. In this work, it is  assumed that the production of particles from a single wounded source conforms to the Negative Binomial Distribution (NBD):
\begin{equation}
    P(n,\overline{n},k) = \frac{\Gamma(n+k)}{\Gamma(n+1)\Gamma(k)}\left(\frac{\overline{n}}{k}\right)^{n}\left(1+\frac{\overline{n}}{k}\right)^{-n-k} \label{eq:NBD}
\end{equation}
In this parametrization of NBD, $\overline{n}$ is the average multiplicity and $1/k$ measures the deviation from a Poisson distribution.
\end{enumerate}

The wounded quark model is an extension of the wounded nucleon approach, where quark-quark collisions instead of nucleon-nucleon  determine particle emission, Refs~\cite{Bialas:2004su, WQM, WCBialas:2012dc}. The WQM holds the main assumptions listed in points above, taking into account that the given description and characterization of the quantities, such as fragmentation function $F(\eta)$ and forward (backward)-going wounded sources $\langle w_{F(B)}\rangle$  refers to the properties of wounded quarks. 
Since the WQM was first proposed, it has been applied 
to study a wide variety of phenomena for various colliding systems and energies, Refs.~\cite{Eremin:2003qn, KumarNetrakanti:2004ym, Fialkowski:2010tr,PHENIX:2013ehw,Mishra:2013dwa,PHENIX:2015tbb,Lacey:2016hqy,Mitchell:2016jio,Zheng:2016nxx,Bozek:2016kpf,Bozek:2017elk,Barej:2017kcw}
\newline
\subsection{Analytical formula}
Determining the analytical formula for the FB strongly intensive quantity $\Sigma$ in the wounded nucleon model is based on the methodology proposed in the papers Refs.~\cite{Bzdak:2009dr, Bzdak:2009xq}. The procedure is discussed in detail  in Appendix~\ref{app:generating}.

The Eq.~\ref{eq:wideeq_sigma} shows the $\Sigma$ formula derived in the WNM framework. 
\begin{widetext}
\begin{equation}
\Sigma=1+\frac{\frac{\overline{n}}{2}C^2\left[ 4k \left({\langle w_{B}\rangle}^2\langle w_{F}^{2}\rangle+\langle w_{B}^{2}\rangle{\langle w_{F}\rangle}^{2} -2\langle w_{F}\rangle \langle w_{B}\rangle\langle w_{F}w_{B}\rangle\right) +8\langle w_{B}\rangle\langle w_{F}\rangle \left( \langle w_{B}\rangle+\langle w_{F}\rangle \right)\right]}{k(\langle w_{F}\rangle+\langle w_{B}\rangle) \left[{(\langle w_{F}\rangle+\langle w_{B}\rangle)}^2 -C^{2}{\left( \langle w_{B}\rangle- \langle w_{F}\rangle\right)}^{2}\right]} \text{,}
 \label{eq:wideeq_sigma}
\end{equation}
\text{where}
\begin{equation*} \label{C_def}
 C=2p-1
\end{equation*}
\end{widetext}
The  parameter $p$ in the formula is the probability at which a wounded nucleon emits a particle to a given pseudorapidity interval. This probability  depends on the position and width of the intervals of interest. The probability $p$ that the particle originating from the wounded nucleon in a forward (backward)-moving nuclei goes into the forward (backward) $\eta$ bin can be determined from Eq.~\ref{eq:probability}.

 \begin{equation}
p=\frac{\int_{F(B)} F(\eta) \,d\eta }{\int_{B} F(\eta) \,d\eta  +\int_{F} F(\eta) \,d\eta }
 \label{eq:probability}
\end{equation}
The probability particle from a forward-moving nucleon  ends up in the backward interval, or vice versa, is $1-p$.

From  Eq.~\ref{eq:wideeq_sigma},  it is immediately  apparent that in the wounded nucleon model:
\begin{itemize}
\item[--] 
 $\Sigma$ values explicitly depend on the number of wounded nucleons, therefore, on the centrality of the collision;
\item[--]
the clear dependence of $\Sigma$ on the number of wounded nucleons seen in the Eq.~\ref{eq:wideeq_sigma} formula implies that $\Sigma$ is no longer a strongly intensive variable;
\item[--]the  $\Sigma$ variable is  sensitive to the parameter $C$ and thus to the probability $p$;
\item[--] the quantity $\Sigma$ is larger than unity ($\Sigma>1$) for  $p\neq 0.5$;
\item[--]  $\Sigma$ is equal to unity only when $C=0$, namely in the case when a wounded nucleon produces particles in F and B pseudorapidity intervals with the  same probability $p=0.5$;

\item[--]  for symmetric collisions $\langle w_F\rangle=\langle w_B\rangle$, Eq.~\ref{eq:wideeq_sigma} can be reduced to the expression:
\begin{equation}
\Sigma=1+\frac{\overline{n}}{2}C^2\left[ \frac{\langle (w_{B} -w_{F})^{2}\rangle}{2\langle w_F\rangle} + \frac{2}{k}\right]
 \label{eq:wideeq_sigma_symetric}
\end{equation}
\end{itemize}
 For the wounded quark model, those analytical formulas  remain valid provided that symbols $w_{F}$ and $w_{B}$ denote the number of forward- and backward-moving wounded quarks. 

One remark is in order here. At first sight, the collapse of the strongly intensive properties of the $\Sigma$ variable through its explicit dependence on the number of wounded constituents observed in the wounded nucleon and wounded quark models seems contradictory since both theoretical frameworks belong to the class of independent source models. However, this originates from the fact that both models formulated above assume, in contrast to  Ref.~\cite{Gorenstein:2011vq},  \emph{two} types of wounded  sources moving forward and backward.
In case the number of wounded constituents moving forward and backward is the same ($w_{F}=w_{B}$), or both sources have the same characteristics, i.e. produce particles with $p=0.5$, the $\Sigma$ variable again regains the strongly intensive properties described in Ref.~\cite{Gorenstein:2011vq}. 

\subsection{MC simulations} \label{s:MCsim}
Information about the Pb--Pb or Xe--Xe collision geometry, such as the wounded constituents' distribution characteristics, was obtained using the Monte Carlo Glauber-like simulation. In this paper, the MC model considers two scenarios for wounded nucleons and wounded quarks.
The simulation can be summarized in three main steps.
\\
\paragraph{\bf Modeling of the nuclei:}
In the simulation, the positions of nucleons and quarks in the nucleus are determined randomly. The location  of nucleons in the nucleus is described by a nuclear density profile modeled by a Woods-Saxon  spherically-symmetric density function:
\begin{equation}
 \rho(\vec{r}) = \frac{\rho_{0}}{1+e^{[(r-R)/a]}}
 \end{equation}
where  $R$ is the nuclear radius, $a$ is the skin depth, and $r=|\vec{r}|$ is the distance from the nucleus center. For ${}^{129}$Xe isotope parameters were adopted to be  $R =5.36 $ $a=0.59$. For ${}^{208}$Pb nucleus,  the sum of individual Woods-Saxon distributions for the proton and the neutron was taken with the following parameters: $R_{p} =6.68 $ and $a_{p}= 0.447$,  $R_{n} =6.69 $ and $a_{n} = 0.570$. 
The parameters' values were adjusted to match those used by the ALICE Collaboration, as described in the Refs.~\cite{XeXecent,centralityALICE:2018tvk}.

 For each nucleon, the distribution of three constituent quarks inside the nucleon is given by the expression:
\begin{equation}
\rho(\vec{r}) = \frac{\rho_{0}}{exp{[-r/a]}} \label{dis_quarks}
 \end{equation}
In formula Eq.~\ref{dis_quarks}, $a = r_{p}/\sqrt{12}$ where $r_{p}$ = 0.81 fm is the proton’s radius, following the Refs.~\cite{Barej:2017kcw, Hofstadter}.
\\
\paragraph{\bf Modeling of the collision process:}
 Constituents (nucleons or quarks) move on straight-line paths and interact according to the inelastic constituent cross-section $\sigma_{ii}^{inel}$.  The value of the cross-section does not depend on the number of individual constituent-constituent collisions. 

 The inelastic nucleon-nucleon cross-section $\sigma_{NN}^{inel}$ was estimated  based on interpolation of pp value at the corresponding center-of-mass-energy. 
In this work, $\sigma_{NN}^{inel}$  values were adapted to correspond to the values used in ALICE Ref.~\cite{centralityALICE:2018tvk} and are: $\sigma_{NN}^{inel} =61.8$ mb at $\sqrt{s_{\rm{NN}}}$= 2.76 TeV, $\sigma_{NN}^{inel} =67.6$ mb at $\sqrt{s_{\rm{NN}}}$= 5.02 TeV and $\sigma_{NN}^{inel} =68.4$ mb at $\sqrt{s_{\rm{NN}}}$= 5.44 TeV.

For the wounded quark model,  to estimate the inelastic quark-quark cross-section, the technique proposed in Ref.~\cite{Barej:2017kcw} was applied. This involved seeking a value of $\sigma_{qq}^{inel}$  for which the  relation
$\sigma_{NN}^{inel}=\int_{0}^{2\pi}\,d\varphi \int_{0}^{+\infty} P(b)b\,db$ is satisfied, for a given energy of the  elementary pp collision. In the latter formula, $ P(b)$ is a probability of proton-proton collision with the impact parameter $b$. Based on the described approach,  the given values were determined: $\sigma_{qq}^{inel} \approx 13.9$ mb at $\sqrt{s_{\rm{NN}}}$= 2.76 TeV, $\sigma_{qq}^{inel} \approx 16.2$ mb at $\sqrt{s_{\rm{NN}}}$= 5.02 TeV and $\sigma_{qq}^{inel} \approx 16.5$ mb at $\sqrt{s_{\rm{NN}}}$= 5.44 TeV.

A nucleon is considered  wounded if it overlaps with the nucleons of the other nucleus in a region of transverse distance radius $d_{NN}\leq\sqrt{\frac{\sigma_{NN}^{inel}}{\pi}}$. Analogically a quark is considered  wounded if it overlaps with the quarks of the other nucleus in a region of transverse distance radius $d_{qq}\leq\sqrt{\frac{\sigma_{qq}^{inel}}{\pi}}$.
\\
\paragraph{\bf Modeling of particle production:}
The model assumes every wounded constituent emits particles with NBD parametrized according to Eq.~\ref{eq:NBD}.  This study considered the production of particles in two kinematic pseudorapidity regions, midrapidity and the forward region.

\subparagraph{Midrapidity:}
 Particle production from a single wounded nucleon in midrapidity was modeled for a sum of forward and backward $\eta$ intervals (F+B) of a width of $\delta\eta=0.2$ each. Values of $\overline{n}$ and $k$ parameters for energies $\sqrt{s_{\rm{NN}}}=$ 2.76, 5.02, and 5.44 TeV were interpolated based on the experimental pp data given in the papers Refs.~\cite{ppALICE:2015olq, ppGhosh:2012xh}, and they are listed in Table~\ref{tab:table_NBD_par1}. The uncertainties of $\overline{n}$ and $k$ were conservatively estimated to be $7\%$ and $10\%$, respectively.
\begin{table}[h]
\caption{\label{tab:table_NBD_par1}%
Values of  NBD parameters  $\overline{n}$ and $k$ (WNM) and $\overline{n_q}$ and $k_q$ (WQM) for particle production in the midrapidity region.
}
\begin{ruledtabular}
\begin{tabular}{c|cc|cc}
$\sqrt{s_{\rm{NN}}}$& $\overline{n}$&$k$& $\overline{n_{q}}=\dfrac{\overline{n}}{\langle N_{q}\rangle}$ &$k_{q}=\dfrac{k}{\langle N_{q}\rangle}$\\
\colrule
2.76 & 1.94 & 1.20 & 1.94/1.50 &1.20/1.50\\
5.02 & 2.16 & 1.14 & 2.16/1.60 &1.14/1.60\\
5.44 & 2.19 & 1.13 & 2.19/1.60&1.13/1.60\\
\end{tabular}
\end{ruledtabular}
\end{table}

  For WQM, values of  NBD parameters were determined by scaling $\overline{n}$ and $k$ by the average number of wounded quarks per wounded nucleon in an elementary collision for a given energy, namely $\overline{n_{q}} =\overline{n}/\langle N_{q}\rangle$ and $k_{q}=k/\langle N_{q}\rangle$. From a simple Monte Carlo calculation, one can determine that these scaling factors are  $\langle N_{q}\rangle\approx1.5$ for  pp collisions at $\sqrt{s}$=2.76 TeV and $\langle N_{q}\rangle\approx1.6$  for  pp collisions at $\sqrt{s}$= 5.02 and 5.44 TeV. 

Particles emitted in the midrapidity region are assigned randomly to either the forward or backward pseudorapidity interval with the probability $p$ and $1-p$, respectively, if they originated from a forward-moving constituent or with the probability $1-p$ and $p$  if they originated from the backward-moving source.

\subparagraph{Forward $\eta$ region:} The particle generation in forward regions was modeled  to  mimic the centrality determination method provided by the V0 ALICE detector. The V0 centrality determination is based on  the sum of charged particles' multiplicity measured in the pseudorapidity ranges 2.8 $<\eta<$ 5.1 and -3.7$<\eta<$-1.7. In this work, the particle generated from a wounded nucleon into the forward region follows the NBD with parameters adopted from the Glauber-NBD fit to V0 detector amplitudes published by ALICE in Ref~\cite{centralityALICE:2018tvk}.
 Again for WQM, the NBD parameters were obtained by scaling $\overline{n}$ and $k$ by the  factor $\langle N_{q}\rangle$ for a given energy. Values of NBD parameters for particle production in the forward region  are listed in Table~\ref{tab:table_NBD_par2}.
\begin{table}[h]
\caption{\label{tab:table_NBD_par2}%
Values of NBD  parameters $\overline{n}$ and $k$ (WNM) and $\overline{n_q}$ and $k_q$ (WQM) for  particle production in the forward $\eta$ region.
}
\begin{ruledtabular}
\begin{tabular}{c|cc|cc}
$\sqrt{s_{\rm{NN}}}$& $\overline{n}$&$k$& $\overline{n_{q}}$&$k_{q}$\\
\colrule
2.76 & 28.7 & 1.6 & 28.7/1.5 &1.6/1.5\\
5.02 & 46.4 & 1.5 & 46.4/1.6 &1.5/1.6\\
5.44 & 53.7 & 1.0 & 53.7/1.6&1.0/1.6\\
\end{tabular}
\end{ruledtabular}
\end{table}

The simulated data samples of Pb--Pb  and Xe--Xe collisions  were divided into centrality classes 
according to cuts on impact parameter $b$ and the number of particles in forward $\eta$ region, $N_{ch}$.  Table~\ref{tab:table_GeometricalProp} provides an overview of the mean number of wounded nucleons $\langle w\rangle = \langle w_{F}\rangle +\langle w_{B}\rangle$  and quarks $\langle w^{q}\rangle= \langle w^{q}_{F}\rangle +\langle w^{q}_{B}\rangle$ calculated for Pb--Pb  and Xe--Xe collisions for various centrality classes defined either by impact parameter $b$ cut or via the number of particles $N_{ch}$. The geometrical properties reported in the table are in good agreement  with the values published by ALICE~\cite{centralityALICE:2018tvk}.
\begin{table}[hbt!]
\caption{\label{tab:table_GeometricalProp}
The mean number of wounded nucleons $\langle w\rangle$ (WNM) and wounded quarks $\langle w_{q}\rangle$ for different centrality classes defined by the cuts on impact parameter $b$ and by the number of particles in the forward $\eta$ region $N_{ch}$. 
}
\begin{ruledtabular}
\begin{tabular}{c|cc|cc}
\multicolumn{5}{c}{}\\
\multicolumn{5}{c}{Pb--Pb collisions at $\sqrt{s_{\rm{NN}}}$= 2.76 TeV}\\
\multicolumn{5}{c}{}\\
\hline
Centrality&\multicolumn{2}{c|}{impact parameter $b$ } &\multicolumn{2}{c}{particle multiplicity $N_{ch}$ }\\
 &$\langle w\rangle$& $\langle w^{q}\rangle$ &$\langle w\rangle$&$\langle w^{q}\rangle$  \\
\hline
$0-10\%$& 354.88   & 978.04  & 354.88 & 979.48  \\
$10-20\%$& 258.90 & 674.51  & 259.69 & 676.31 \\
$20-30\%$& 185.39 & 458.82   & 185.98 & 460.04   \\
$30-40\%$& 128.83 & 300.65   &129.21 & 301.32  \\
$40-50\%$& 85.66  & 186.13  & 85.89 & 186.35 \\
$50-60\%$& 53.42  & 106.57  & 53.57 & 106.38  \\
$60-70\%$& 30.64  & 55.36   & 30.60 & 54.77 \\
$70-80\%$& 15.79  & 25.92  & 15.53 & 24.98 \\
\hline
\multicolumn{5}{c}{}\\
\multicolumn{5}{c}{Pb--Pb collisions at $\sqrt{s_{\rm{NN}}}$= 5.02 TeV}\\
\multicolumn{5}{c}{}\\
\hline
Centrality&\multicolumn{2}{c|}{impact parameter $b$ } &\multicolumn{2}{c}{particle multiplicity $N_{ch}$ }\\
 &$\langle w\rangle$& $\langle w^{q}\rangle$ &$\langle w\rangle$&$\langle w^{q}\rangle$  \\
\hline
$0-10\%$& 357.66  & 993.20  & 357.60 & 994.51 \\
$10-20\%$& 262.18 & 689.14  & 263.00 & 691.05 \\
$20-30\%$& 188.12 & 470.68   & 188.70 & 471.90  \\
$30-40\%$& 131.03 & 309.78   & 131.47 & 310.55 \\
$40-50\%$& 87.22  & 192.32  & 87.49 & 192.57 \\
$50-60\%$& 54.54  & 110.61  & 54.63 & 110.36 \\
$60-70\%$& 31.26  & 57.46   & 31.21 & 56.87 \\
$70-80\%$& 16.07  & 26.81  & 15.83 & 25.86 \\
\hline
\multicolumn{5}{c}{}\\
\multicolumn{5}{c}{Xe--Xe collisions at $\sqrt{s_{\rm{NN}}}$= 5.44 TeV}\\
\multicolumn{5}{c}{}\\
\hline
Centrality&\multicolumn{2}{c|}{impact parameter $b$ } &\multicolumn{2}{c}{particle multiplicity $N_{ch}$ }\\
 &$\langle w\rangle$& $\langle w^{q}\rangle$ &$\langle w\rangle$&$\langle w^{q}\rangle$  \\
\hline
$0-10\%$& 221.91  & 605.79  & 221.32  & 606.56  \\
$10-20\%$& 164.56 & 420.25  & 165.58 & 422.44 \\
$20-30\%$& 117.89 & 283.89   & 118.65 & 285.24   \\
$30-40\%$& 81.75 & 184.48   & 82.30 & 185.39  \\
$40-50\%$& 54.20  & 113.62  & 54.51 & 113.88 \\
$50-60\%$& 33.82  & 65.53  & 34.00 & 65.35  \\
$60-70\%$& 19.78  & 35.37   & 19.74 & 34.80 \\
$70-80\%$& 11.00  & 18.31  & 10.54 & 17.12 \\
\end{tabular}
\end{ruledtabular}
\end{table}

The distributions of wounded nucleons and quarks obtained as described in this section were used to calculate the $\Sigma$ in the wounded nucleon and wounded quark models based on the analytical expression Eq.~\ref{eq:wideeq_sigma}. Additionally, numerical Monte Carlo calculations, which incorporate particle production from a wounded constituent, were used to calculate $\Sigma$  in the discussed models based on the definition of this variable, Eq.~\ref{sigma_def}.

\section{Results and Discussion}
\subsection{Centrality dependence of the forward-backward quantity $\Sigma$ in WNM and WQM}\label{s:R1}

\begin{figure*}
\includegraphics[width=0.95\textwidth]{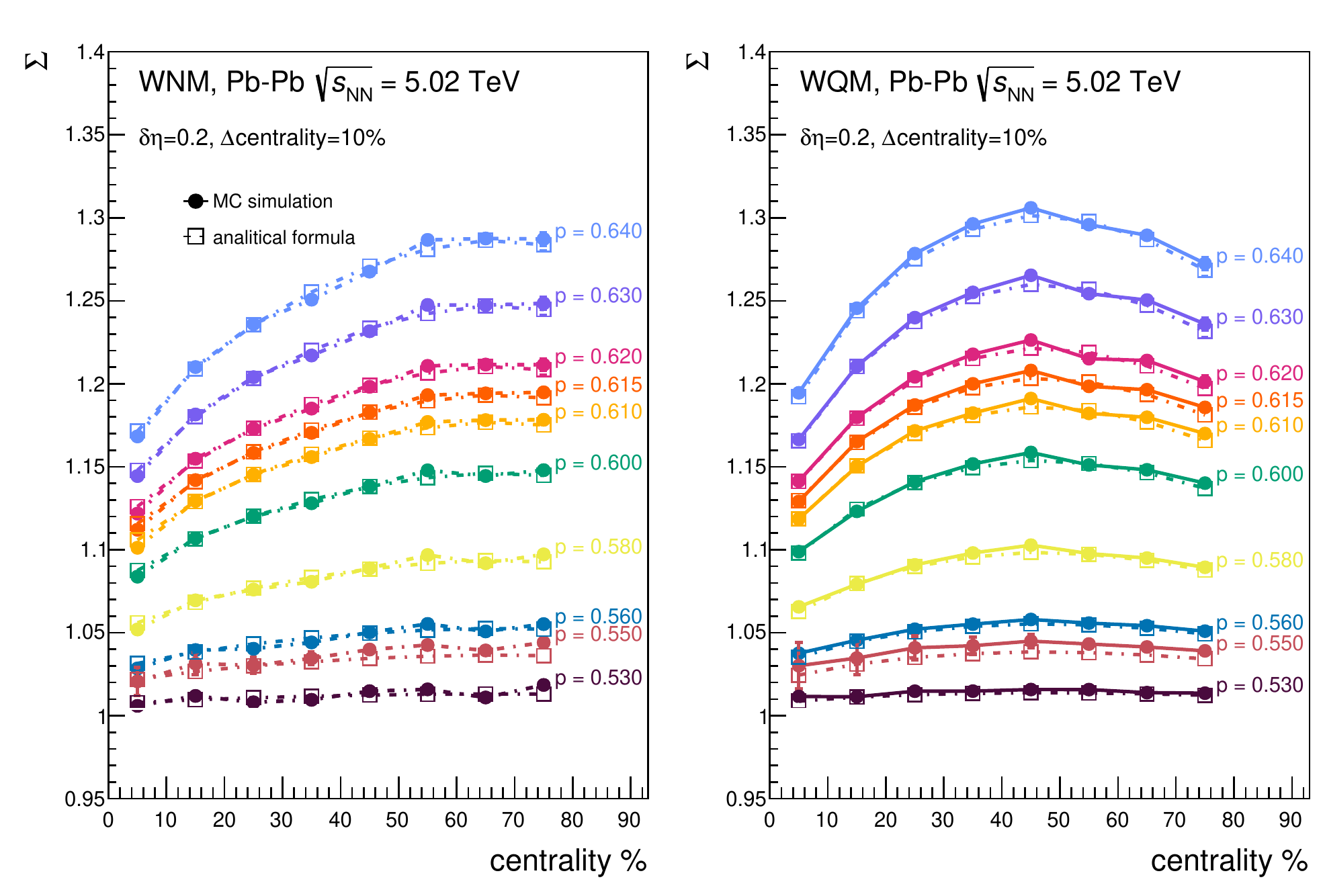}
\caption{\label{fig:Sigma_analytical_vs_MCsim} The $\Sigma$ observable obtained for (a) wounded nucleon and (b) wounded quark model-generated Pb--Pb collisions at $\sqrt{s_{\rm{NN}}}$= 5.02 TeV as a function of centrality class. Results
are presented for several values of the probability $p$. Statistical uncertainties are labeled with vertical bars. Centrality selection is made via impact parameter $b$.}
\end{figure*}

Figure~\ref{fig:Sigma_analytical_vs_MCsim} presents the overview of the results on forward-backward correlations with $\Sigma$ obtained in wounded nucleon and wounded quark models in Pb--Pb collisions at $\sqrt{s_{\rm{NN}}}$= 5.02 TeV as a function of centrality class. Results were determined for $10\%$ width of centrality classes of  ranges from  $0-10\%$, $10-20\%$, $20-30\%$, etc., up to peripheral collisions  $70-80\%$. The results are presented for selected probability values varying from $p=0.5$ to $p=0.64$. The figure compares the $\Sigma$ quantity derived from the analytical formula Eq.~\ref{eq:wideeq_sigma} and  that obtained from the definitions of this variable given by Eq.~\ref{sigma_def} in the numerical Monte Carlo simulation. 

Based on the presented in Fig.~\ref{fig:Sigma_analytical_vs_MCsim} results, one can  list the following observations.
\begin{itemize}
    \item [--] There is  full agreement, within statistical uncertainties, between the values of $\Sigma$ obtained from the analytical formula and the numerical MC calculations that include particle production for all studied centrality classes. This is valid for both the wounded nucleon and wounded quark models. Consequently, the WNM and WQM results in the following sections will only be determined by the analytical formula.
    \item [--] The values of $\Sigma$ quantity grow with the  values of probability  $p$. This increase is a bit steeper for WQM. 
     \item [--] For $p=0.5$, the $\Sigma$ is equal to unity for all centrality classes as predicted by the Eqs.~\ref{eq:wideeq_sigma} and~\ref{eq:wideeq_sigma_symetric} for both  wounded constituent models. 
    \item [--] For $p\neq0.5$ the results for $\Sigma$ tend to increase  from central   to peripheral  collisions. This increase is monotonic for WNM.   For WQM, the $\Sigma$ values take a maximum around the mid-central collisions $40-50\%$ and then  gently fall towards the most peripheral sample. This centrality dependence of $\Sigma$ quantity is more evident the higher the value of probability $p$. 
    \item [--] The most surprising aspect of the presented results  is that both WNM and WQM seem to resemble the behaviour observed in the  experimental data Ref.~\cite{Sputowska:2022gai}, particularly for $p\sim $ 0.6. What is striking, wounded nucleon and wounded quark models are able to describe the $\Sigma$ trend with centrality much better than the most  widely used heavy-ion  models such as HIJING, AMPT, or EPOS (see Ref.~\cite{Sputowska:2022gai} for more details).
    \item [--] The values of $\Sigma$ appear to change with  the size of the system (centrality). This  is not caused by differences in the characteristics of the particle-producing source but rather by the fact that $\Sigma$ is \emph{not} strongly intensive in WNM and WQM, and it can be affected by the volume of the system.  This topic will be discussed  in more detail in the next sections.
\end{itemize}

\subsection{ The $\Sigma$ quantity as a function of centrality bin width and centrality selection method in WNM and WQM}\label{s:R2}

From the expressions Eqs.~\ref{eq:wideeq_sigma} and \ref{eq:wideeq_sigma_symetric},  it is clear that in the  wounded constituent approach,   $\Sigma$ exhibits an intrinsic and non-trivial dependence on the number of forward- and backward-moving wounded sources, hence it is not anymore a strongly intensive quantity. Thus, it was found particularly interesting to verify how the volume fluctuations and selection of centrality estimator affect the $\Sigma$ quantity.

\begin{figure*}
\includegraphics[width=0.95\textwidth, trim={0.1cm 0.1cm 0.1cm 0.1cm},clip]{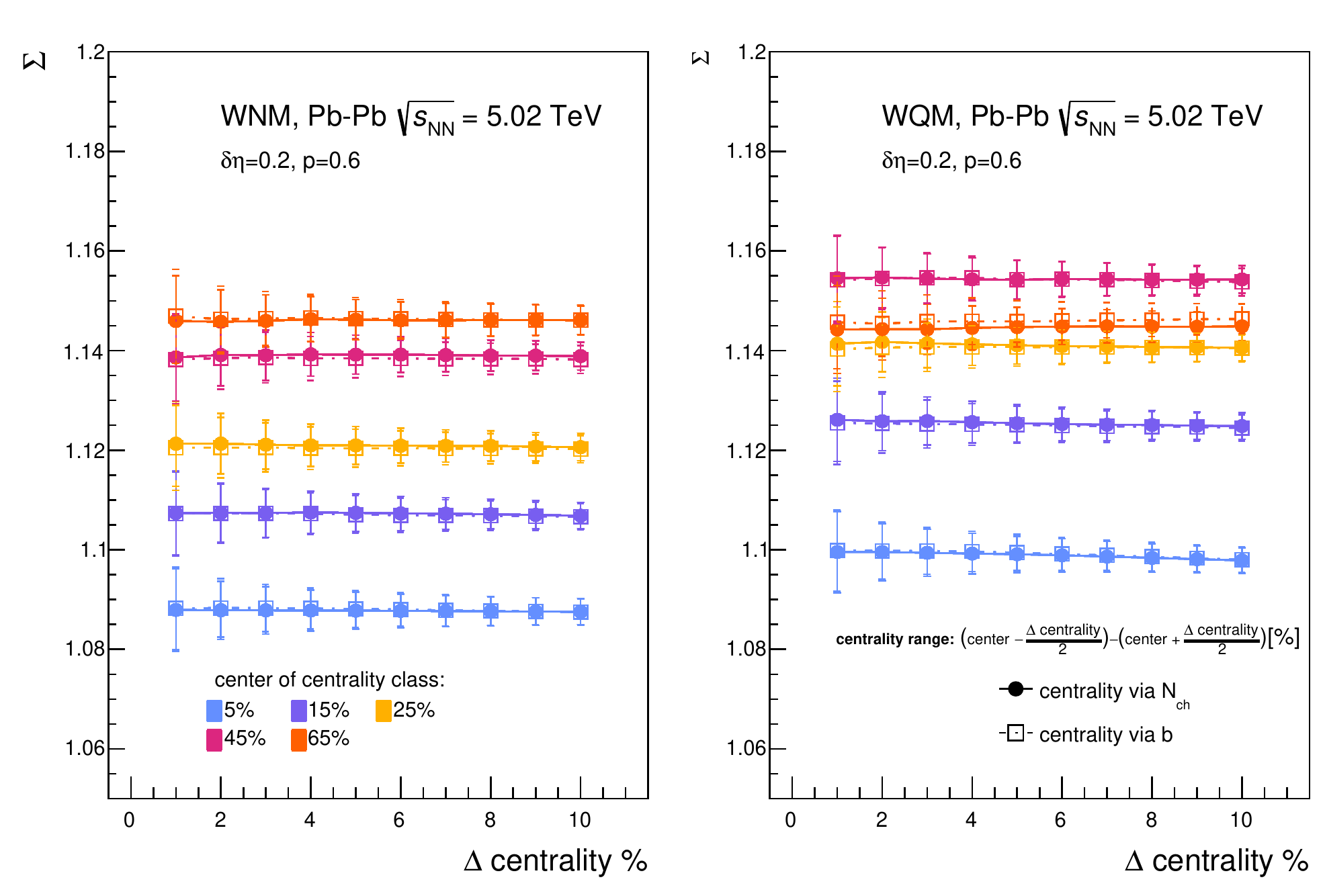}
\caption{\label{fig:WNM_cent_sigma} The $\Sigma$ observable obtained for (a) wounded nucleon and (b) wounded quark model-generated Pb–Pb
collisions at $\sqrt{s_{\rm NN}}$ = 5.02 TeV, drawn as a function of centrality class size ($\Delta \text{ centrality}$), for a fixed
value of the probability $p =0.6$. The results are obtained for different centrality selection methods: via impact parameter $b$ and via particle multiplicity $N_{ch}$. The width of the centrality class ($\Delta \text{ centrality}\leq10\%$) changes from $1\%$ to $10\%$. The ranges of centrality class for a given centrality bin width can be deduced from the formula in the figure. The different colors of the data points correspond to different centralities of the Pb–Pb collision. Statistical uncertainties are labeled with vertical bars. }
\end{figure*}

Figure~\ref{fig:WNM_cent_sigma} shows the dependence of $\Sigma$ quantity on the width of centrality class ($\Delta \text{ centrality}$) for Pb--Pb collisions at $\sqrt{s_{\rm NN}}$=5.02 TeV in the wounded nucleon  and wounded quark model, respectively.  This study was carried out for different centrality classes of Pb–-Pb collisions, from central to peripheral. The width of the centrality class varied from its maximum range of $10\%$, where the largest contribution from volume fluctuations is expected, down to $1\%$ centrality bin width. Results are presented for two  different centrality selection criteria via impact parameter $b$ and particle multiplicity $N_{ch}$. 

From the data in Fig.~\ref{fig:WNM_cent_sigma}, it is apparent that $\Sigma$ in the framework of  WNM and WQM in symmetric collisions such as Pb--Pb  does not depend on centrality bin width in studied centrality classes. This implies it is insensitive to volume fluctuation.  The comparison between results obtained for two centrality selection methods revealed that the $\Sigma$ observable also does not depend on  the way centrality is determined (also shown in the figure).

The lack of dependence of $\Sigma$ on the width of the centrality interval and the method of centrality selection noted for $\Delta \text{ centrality}\leq10\%$ in the wounded constituent framework is rather surprising, given that the $\Sigma$ quantity is explicitly dependent on the term related to the fluctuations of wounded constituents. These apparent "strongly-intensive-quantity-like"  properties of $\Sigma$ for centrality classes of widths $\Delta \text{ centrality}\leq10\%$ in symmetric nucleus-nucleus collisions resemble the behavior reported by ALICE experimental data Refs.~\cite{Sputowska:2019yvr, Sputowska:2022CPOD, Sputowska:2022gai}, and can be explained theoretically  if one notes that the Eq.~\ref{eq:wideeq_sigma_symetric} can be rewritten in terms of \emph{partial covariance}.

 \subsection{ The relation between the $\Sigma$ quantity and  partial covariance in WNM and WQM for fully symmetric collision}\label{s:R3}
The partial covariance is defined as: 
\begin{equation}
 \text{Cov}(X,Y \bullet Z) = \text{Cov}(X,Y) -\frac{\text{Cov}(X,Z) \text{Cov}(Y,Z) }{\text{Var}(Z)}\label{partial_cov}
\end{equation}
and measures the degree of relation between two random variables $X$ and $Y$, with the effect of controlling random variable $Z$ removed. 
An exhaustive description of the partial correlations technique in the domain of heavy-ion physics is given in Ref.~\cite{Olszewski:2017vyg}. As the authors of  the mentioned  paper explain in more detail, for a sample that  satisfies the affine condition $E(X\mid Y)=a+BY$, the partial covariance is related to conditional covariance $\text{Cov}(X,Y \bullet Z) \simeq \text{Cov}(X,Y \mid Z)$.  The latter relation implies that the measurement of the partial covariance with a control variable related to collision geometry, such as the number of wounded nucleons, can be interpreted as a conditional covariance with a fixed  number of wounded nucleons in the control interval.

\begin{figure*}
  \includegraphics[width=1.0\textwidth]{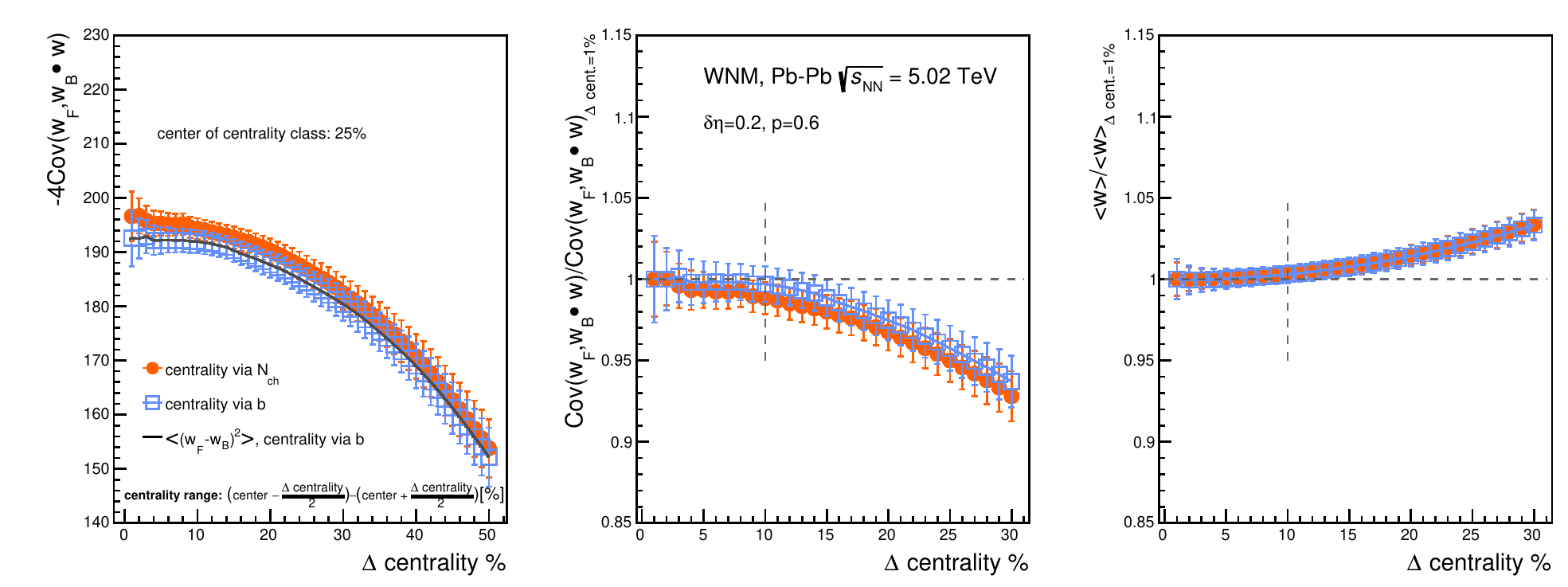}
    \caption{The partial covariance $\text{Cov}(w_{F},w_{B} \bullet w)$ and average number of wounded nucleons $\langle w\rangle$ as a function of the width of centrality class ($\Delta \text{ centrality}$). The center of the centrality class is $25\%$, and the width of the centrality class changes from $\Delta \text{ centrality}$ =1$\%$ up to $\Delta \text{ centrality}$ =50$\%$. 
    The left panel shows the value of partial covariance $\text{Cov}(w_{F},w_{B} \bullet w)$, the middle and right panels display the partial covariance value and the average number of wounded nucleons $\langle w\rangle$, respectively, both normalized to their values determined at $\Delta \text{ centrality}$ =1$\%$. The ranges of centrality class for a given centrality bin width can be deduced from the formula in the figure.}
    \label{fig:partial_cov}
\end{figure*}

It can be noticed that for a symmetric collision, $\text{Var}(w_{F})=\text{Var}(w_{B})$ and $\langle w_{F}\rangle=\langle w_{B}\rangle$,  the following relation occurs:
\begin{equation}\label{FLUC_COV}
\langle (w_{B} -w_{F})^{2}\rangle =-4\text{Cov}(w_{F},w_{B} \bullet w), 
\end{equation}
The right side of the equation is a  partial covariance between the forward and backward-moving wounded nucleons  with the total number of wounded nucleons $w=w_{F} +w_{B}$ as a control variable. The analytical proof that the above expression is valid is presented in the Appendix~\ref{app:cov}; the graphical one is presented in the left panel of Fig.~\ref{fig:partial_cov}, where the quantities determined by the left and right sides of the equation have been directly compared.

Some comment about $\text{Cov}(w_{F},w_{B} \bullet w)$ properties is necessary here. From the discussed earlier characteristics of partial covariance emerges that it is equivalent to the conditional covariance between $w_F$ and $w_B$ at the fixed average number of wound nucleons for a given centrality class\footnote{ The affine condition for wounded nucleons:  $E(w_{F}\mid w_{B})=a+Bw_{B}$ is almost always satisfied, regardless of the sample (centrality class) size.}. Thus, the value of $\text{Cov}(w_{F},w_{B} \bullet w)$ does not change when increasing the centrality interval around its center as long as the value of the average number of wounded nucleons remains unchanged. From Fig.~\ref{fig:partial_cov}, it can be concluded that the average value of the number of wounded nucleons $\langle w\rangle$, determined for a center of centrality class in question, does not 
vary when the width of the centrality class fluctuates no more than $10\%$ around its center.  
Moreover, considering that the average number of wounded nucleons is independent of centrality selection criteria~\cite{Konchakovski:2008cf}, thus values of $\text{Cov}(w_{F},w_{B} \bullet w)$ do not change when  applying different  centrality estimators. In other words, independently of the width of the centrality interval and the method of centrality selection, the term $\text{Cov}(w_{F},w_{B} \bullet w)$ always gives a value of covariance between $w_F$ and $w_B$  when the average number of the wounded nucleons in the studied centrality sample remains fixed.

With the substitution provided by relation Eq.~\ref{FLUC_COV} the Eq.~\ref{eq:wideeq_sigma_symetric} can be rewritten in an equivalent form:
\begin{equation}\label{eq:wideeq_sigma_symetric_partial}
\Sigma=1+\frac{\overline{n}}{2}C^2\left[\frac{-2\text{Cov}(w_{F},w_{B} \bullet w)}{\langle w_F\rangle} + \frac{2}{k}\right].
\end{equation}
This formula explains all the properties of $\Sigma$ in the wounded constituent framework seen in Fig.~\ref{fig:WNM_cent_sigma}. From the formula, one can conclude that:
\begin{itemize}
    \item [--]  for centrality classes of  $\Delta\text{ centrality}\leq10\%$, the $\Sigma$ variable does not depend on the centrality bin width as all of the terms in Eq.~\ref{eq:wideeq_sigma_symetric_partial} are volume fluctuation-free quantities. 
    \item [--]  all terms including $\text{Cov}(w_{F},w_{B} \bullet w)$ and $\langle w_F\rangle$  are  invariant under the change of centrality estimator, which implies that $\Sigma$  does not depend on the criterion of centrality selection.  
     \item [--] since the ratio $\frac{-2\text{Cov}(w_{F},w_{B} \bullet w)}{\langle w_F\rangle}$ depends on the volume of the system and, more precisely, on the average number of wounded nucleons, so does $\Sigma$. 
\end{itemize}

Overall, these results indicate that $\Sigma$ in the wounded nucleon model, even though it does not depend on volume fluctuations nor the method used for centrality estimation, is generally not strongly intensive as it is sensitive to the average number of wounded nucleons (system size) in the sample. However, for relatively narrow centrality ranges,  $\Sigma$ exhibits the properties of a strongly intensive quantity. 

Analogous conclusions can be drawn for the wounded quark model.

\subsection{Comparison between experimental results and wounded constituent models}\label{s:R4}

\begin{figure*}
\includegraphics[width=0.95\textwidth]{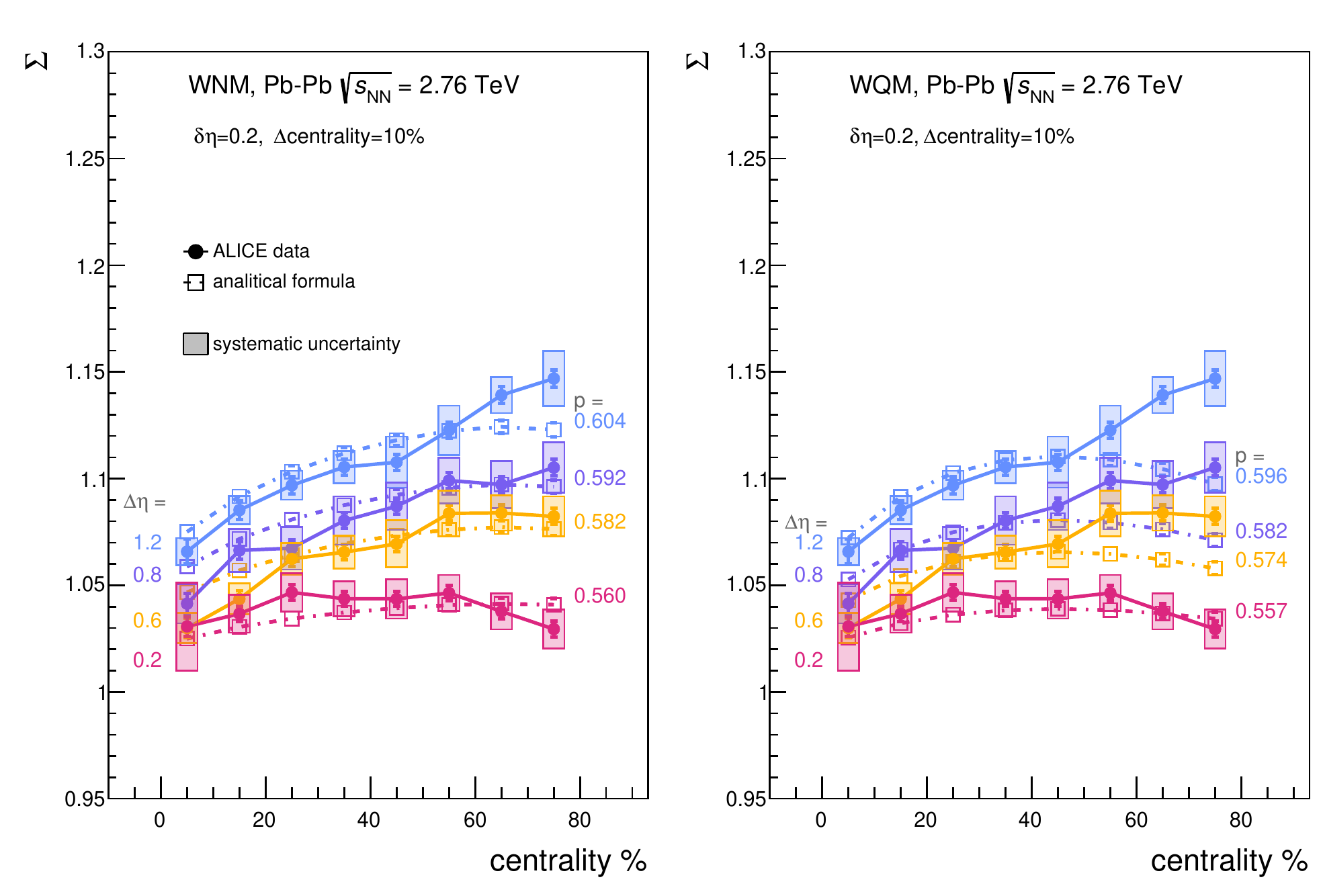}
\caption{\label{fig:SIGMA_analitical_vs_data_data2.76TeV} The centrality dependence of strongly intensive quantity $\Sigma$ obtained in WNM (left panel) and WQM (right panel) for Pb--Pb collisions at $\sqrt{s_{\rm{NN}}}$ = 2.76 TeV. Data points are drawn for selected fixed values of probability  $p$. Theoretical predictions at each panel are compared with ALICE experimental results, which were redrawn from Ref.~\cite{FiguresRes} for several values of the pseudorapidity gap $\Delta\eta=$1.2, 0.8, 0.6, 0.2. For the models, the marked uncertainty bars denote the total uncertainty.}
\end{figure*}

\begin{figure*}
\includegraphics[width=0.95\textwidth]{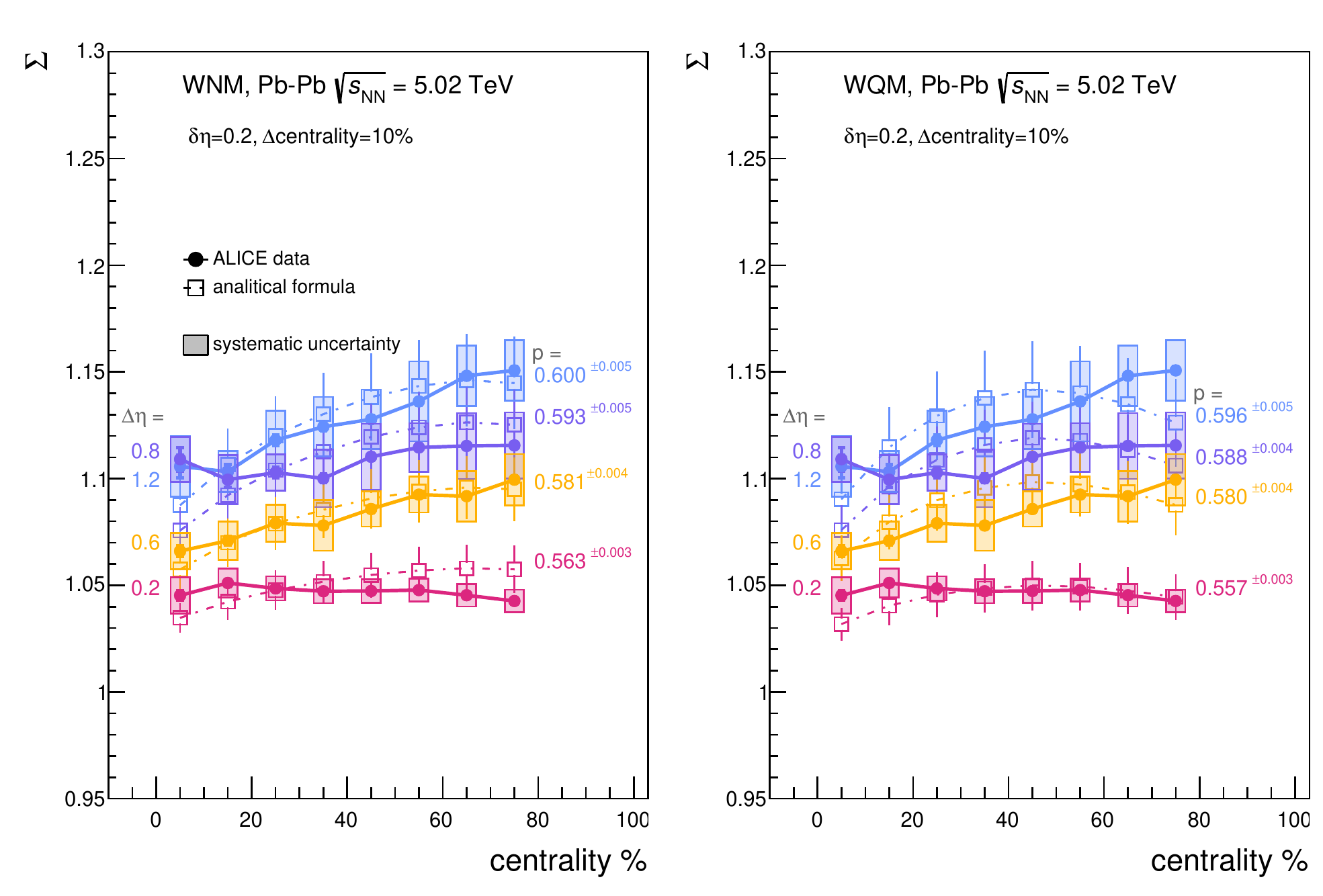}
\caption{\label{fig:SIGMA_analitical_vs_data_data5.02TeV} The centrality dependence of strongly intensive quantity $\Sigma$ obtained in WNM (left panel) and WQM (right panel) for Pb--Pb collisions at $\sqrt{s_{\rm{NN}}}$ = 5.02 TeV. Data points are drawn for selected fixed values of probability  $p$. Theoretical predictions at each panel are compared with ALICE experimental results, which were redrawn from Ref.~\cite{FiguresRes} for several values of the pseudorapidity gap $\Delta\eta=$1.2, 0.8, 0.6, 0.2. For the models, the marked uncertainty bars denote the total uncertainty.}

\end{figure*}
\begin{figure*}
\includegraphics[width=0.95\textwidth]{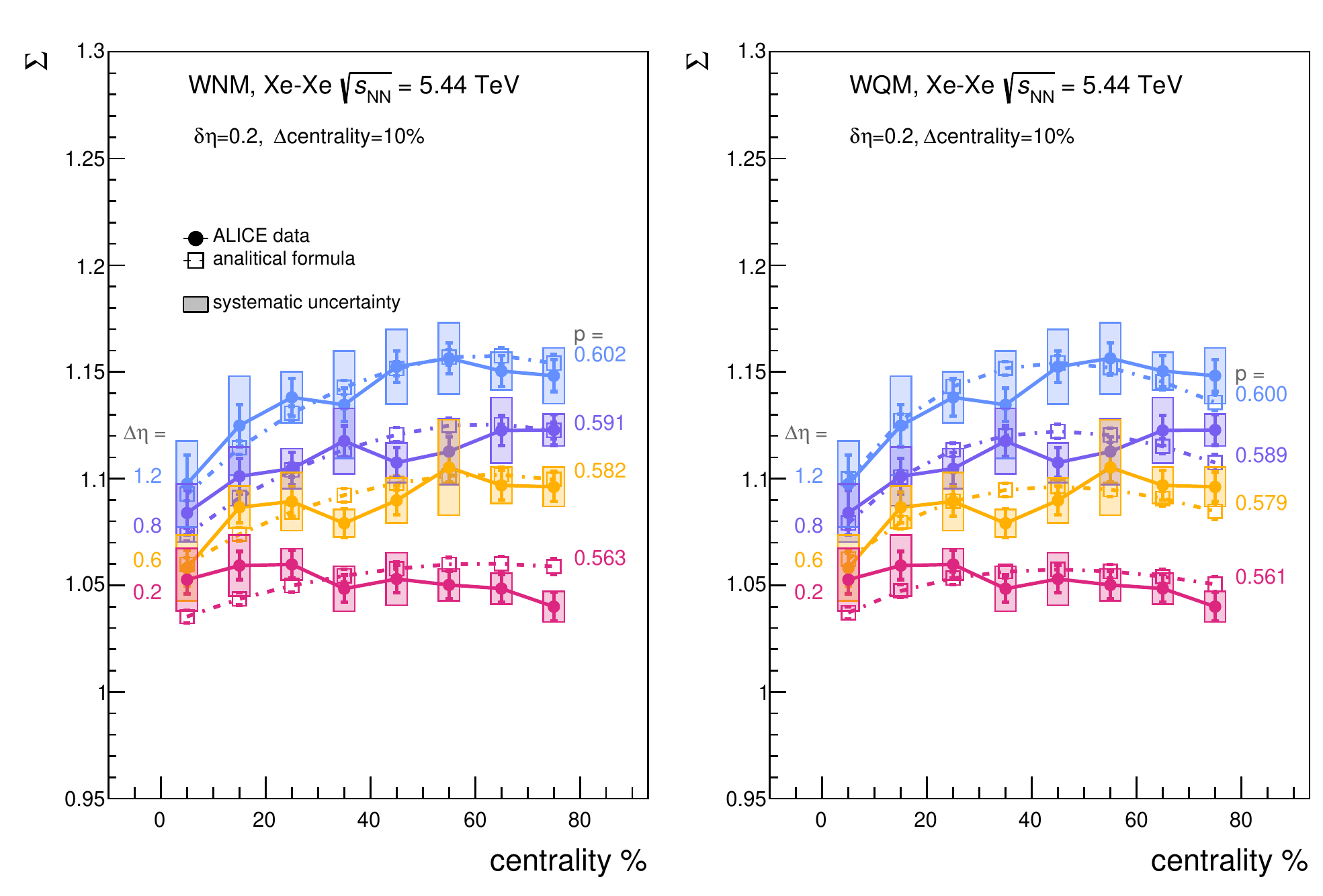}
\caption{\label{fig:SIGMA_analitical_vs_data_XeXe} The centrality dependence of strongly intensive quantity $\Sigma$ obtained in WNM (left panel) and WQM (right panel) for Xe--Xe collisions at $\sqrt{s_{\rm{NN}}}$ = 5.44 TeV. Data points are drawn for selected fixed values of probability  $p$. Theoretical predictions at each panel are compared with ALICE experimental results, which were redrawn from Ref.~\cite{FiguresRes} for several values of the pseudorapidity gap $\Delta\eta=$1.2, 0.8, 0.6, 0.2. For the models, the marked uncertainty bars denote the total uncertainty.}
\end{figure*}

Figures~\ref{fig:SIGMA_analitical_vs_data_data2.76TeV}-\ref{fig:SIGMA_analitical_vs_data_XeXe} present the overview of the results on forward-backward correlation with $\Sigma$ in the wounded nucleon and wounded quark model in Pb--Pb collisions at $\sqrt{s_{\rm{NN}}}$ =2.76 and 5.02 TeV and Xe--Xe collisions at $\sqrt{s_{\rm{NN}}}$ =5.44 TeV. Data points  are plotted as a function of centrality for $\Delta \text{ centrality}$ =10$\%$. In each figure, theoretical predictions are compared to the values of $\Sigma$  measured in the experimental data sample of Pb--Pb and Xe--Xe collisions with the ALICE detector at the LHC. ALICE results were redrawn from Ref.~\cite{FiguresRes}. The experimental data points are presented for several fixed values of the separation gap ($\Delta\eta$) between forward and backward intervals.
 
 The results for the wounded nucleon and wounded quark models are shown for several values of probability $p$, which plays the role of the model parameter. The values of $p$ were estimated in such a way as to provide the best fit of the models to the experimental data. The error bars for WNM and WQM results in the figure represent the total uncertainty, including both the statistical uncertainty and the uncertainty originating  from the variation of  $\overline{n}$ and $k$.

\begin{figure*}
\includegraphics[width=0.99\textwidth]{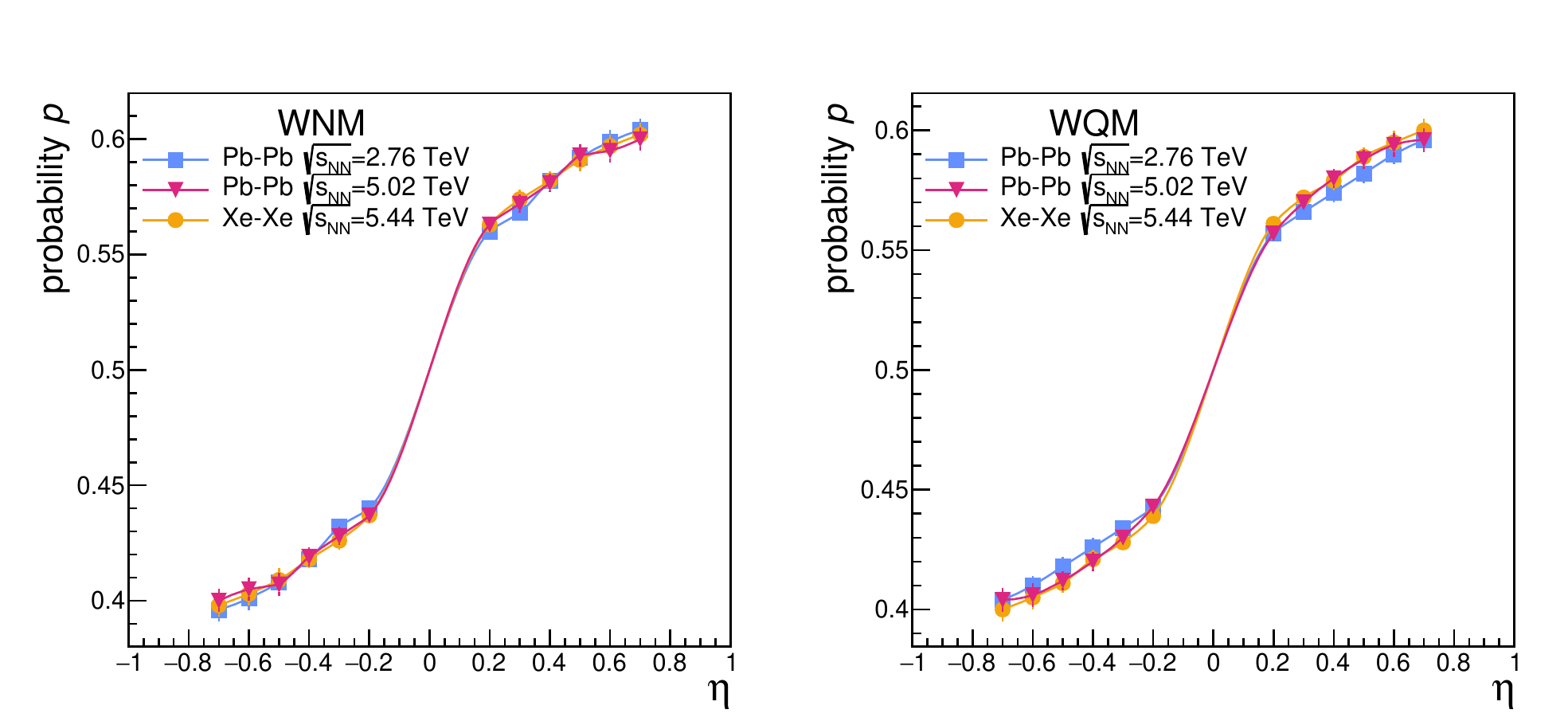}
\caption{\label{fig:prob} Probability $p$ as a function of pseudorapidity  obtained in WNM (left panel) and WQM (right panel) for Pb--Pb and Xe--Xe collisions. Values of $p$ are drawn with total uncertainties, to their values contributed the most systematic uncertainty of experimental data for  $\Sigma$ and uncertainty on  NBD parameters.
}
\end{figure*}

Overall, the findings outlined in Figs.~\ref{fig:SIGMA_analitical_vs_data_data2.76TeV}-\ref{fig:SIGMA_analitical_vs_data_XeXe} indicate the following:
\begin{itemize}
\item[--] The energy dependence of the $\Sigma$ quantity in models closely matches one observed in  the experimental data. When the system energy increases, the $\Sigma$ value also increases, regardless of the system type. According to the formula Eq.~\ref{eq:wideeq_sigma_symetric}, the energy dependence of $\overline{n}$ and $k$ plays a role in this behavior, as an increase in $\overline{n}$ and a decrease in $k$ with collision energy (Table~\ref{tab:table_NBD_par1}) contribute to the noted trend.
\item[--] The models accurately depict the trend of $\Sigma$ values with centrality manifested by the experimental data. The $\Sigma$ values increase from central to peripheral collisions. Eq.~\ref{eq:wideeq_sigma_symetric} shows that this growth is caused by a term associated with wounded nucleons (quarks) since all remaining parts of the $\Sigma$ formula are robust against centrality.
\item[--] The Figs.~\ref{fig:SIGMA_analitical_vs_data_data2.76TeV} and~\ref{fig:SIGMA_analitical_vs_data_data5.02TeV} show that qualitatively, the WNM model reproduces the behavior of Pb--Pb data better than WQM, especially in the peripheral area. 
\item[--]The best agreement between the experimental results on $\Sigma$ and the theoretical description, both WNM and WQM,  is seen for the Xe--Xe collision, Fig.~\ref{fig:SIGMA_analitical_vs_data_XeXe}.
\item[--] As reported earlier in this paper, the values of the $\Sigma$ in the wounded nucleon (quark) model show a sensitivity to the probability value $p$, which is a rather unexpected outcome. WNM and WQM effectively describe the data within the studied kinematic range of $\eta$ when the probability values $p$ exceed 0.5. When comparing the data with the models, it was found that the probability $p$ changes as a function of pseudorapidity. 
\item[--]The relation between probability $p$ and pseudorapidity is illustrated in Fig.~\ref{fig:prob}.  From this figure emerges that within the limits of uncertainty,  $p$ shows a universal trend as a function of pseudorapidity, regardless of the type of colliding system or its energy within the studied energy range.
\end{itemize}

The wounded constituent approach describes the $\Sigma$ observable in overall agreement  with the data and surpasses more complex models like HIJING, AMPT, or EPOS, which fail to describe this observable's behavior properly.
Perhaps one of the most surprising outcomes of this study is a direct relation between probability $p$ and $\Sigma$ in the WNM and WQM. In Pb--Pb and Xe--Xe collisions, reliable probability $p$ values for the wounded constituent  model can be extracted from experimental results on $\Sigma$. Interestingly, these probability values provide a way to estimate the fragmentation function in these symmetric collisions using Eq.~\ref{eq:probability}.  This procedure is discussed  in the next section.

\subsection{Wounded nucleon and wounded quark fragmentation functions} \label{s:R5}
The term fragmentation function $F(\eta)$ of a wounded constituent refers to the pseudorapidity single-particle density characterizing the contribution  from a wounded nucleon or a wounded quark to particle production.
For a two-component picture of forward- and backward-moving wounded constituents, the fragmentation function can be extracted based on the measurement of a single particle pseudorapidity distribution, $N(\eta)$, from the relation Eq.~\ref{fragmenation_function0}.
\begin{equation}\label{fragmenation_function0}
    F(\eta)=\dfrac{1}{2}\left(\dfrac{N(\eta)+N(-\eta)}{\langle w_{F}\rangle+\langle w_{B}\rangle}+\dfrac{N(\eta)-N(-\eta)}{\langle w_{F}\rangle-\langle w_{B}\rangle}\right)
\end{equation}
The  formula above  is a regular transformation of the Eq.~\ref{eq:particle_density},  Ref~\cite{Barej:2017kcw}.

The major problem in this approach is that  Eq.~\ref{fragmenation_function0}  is only applicable for asymmetric collisions for which $w_{F}\neq w_{B}$ since for the symmetric case ($w_{F}=w_{B}$) the second part of this relation becomes indefinite.

One method for determining the fragmentation function in a symmetric pp collision was proposed in Ref.~\cite{Bzdak:2009bc} based on the study of the average numbers of produced particles together with  pairs of particles in different symmetric and asymmetric pseudorapidity intervals.

In this section, it will be shown that measuring the forward-backward correlations with the $\Sigma$ variable provides a unique opportunity to determine the fragmentation function of a wounded nucleon or quark in a  symmetric nucleus-nucleus collision. 
The method is based on the fact that the $\Sigma$ quantity in the wounded constituent model provides direct information about the probability value $p$, which is related to the single nucleon fragmentation function through Eq.~\ref{eq:probability}.
By combining Eq.~\ref{eq:particle_density} and  Eq.~\ref{eq:probability}, it the following relation can be derived:
\begin{eqnarray}
&\int_{F} F(\eta) \,d\eta =p{\int_{F} \left(F(-\eta)   +F(\eta)\right)\,d\eta }\nonumber\\
&=\dfrac{p}{\langle w_{F}\rangle+\langle w_{B}\rangle}{\int_{F} \left(N(-\eta)   +N(\eta)\right)\,d\eta }.
 \label{eq:prob_ff}
\end{eqnarray}
For a sufficiently narrow pseudorapidity ($\eta$) interval F, one can obtain the following:
\begin{eqnarray}
F(\eta)  \approx\dfrac{p}{\langle w_{F}\rangle+\langle w_{B}\rangle} \left(N(-\eta)   +N(\eta)\right) .
 \label{eq:prob_ff}
\end{eqnarray}

The fragmentation functions of the wounded nucleon and wounded quark were determined for minimum bias as well as for different centralities of Pb--Pb collisions at $\sqrt{s_{\rm{NN}}}$=5.02 TeV, according to Eq.~\ref{eq:prob_ff}. The values of the $p$ parameter,  entering the formula,  were extracted based on the selection of the best fit between the $\Sigma$ obtained in the wounded constituent model to the experimental results measured by ALICE, as described in Section~\ref{s:R4}. For $\eta$=0, it was assumed that the probability value $p$=0.5, which is also suggested by the behavior of the data points in Fig.~\ref{fig:prob}. The  $N(\pm\eta)$ values were taken from charged-particle multiplicity densities measured in Pb--Pb collisions at $\sqrt{s_{\rm{NN}}}$=5.02 TeV by ALICE and reported in Ref.~\cite{PbPbALICE:2016fbt}.   The number of wounded constituents ($\langle w_{F}\rangle+\langle w_{B}\rangle$), nucleons and quarks, was generated in Monte Carlo discussed in Section~\ref{s:MCsim}.
This study considered a sample with a $0-80\%$ centrality as a minimum bias since centrality class $80\%$  has been affected by contamination from electromagnetic processes. Given that the publication Ref.~\cite{PbPbALICE:2016fbt} does not directly provide information on $dN_{ch}/d\eta$ for centrality $0-80\%$, the values were determined from the available data as an arithmetic average. Their uncertainties were estimated from the sum in quadrature. 
\begin{figure*}
\includegraphics[width=0.99\textwidth]{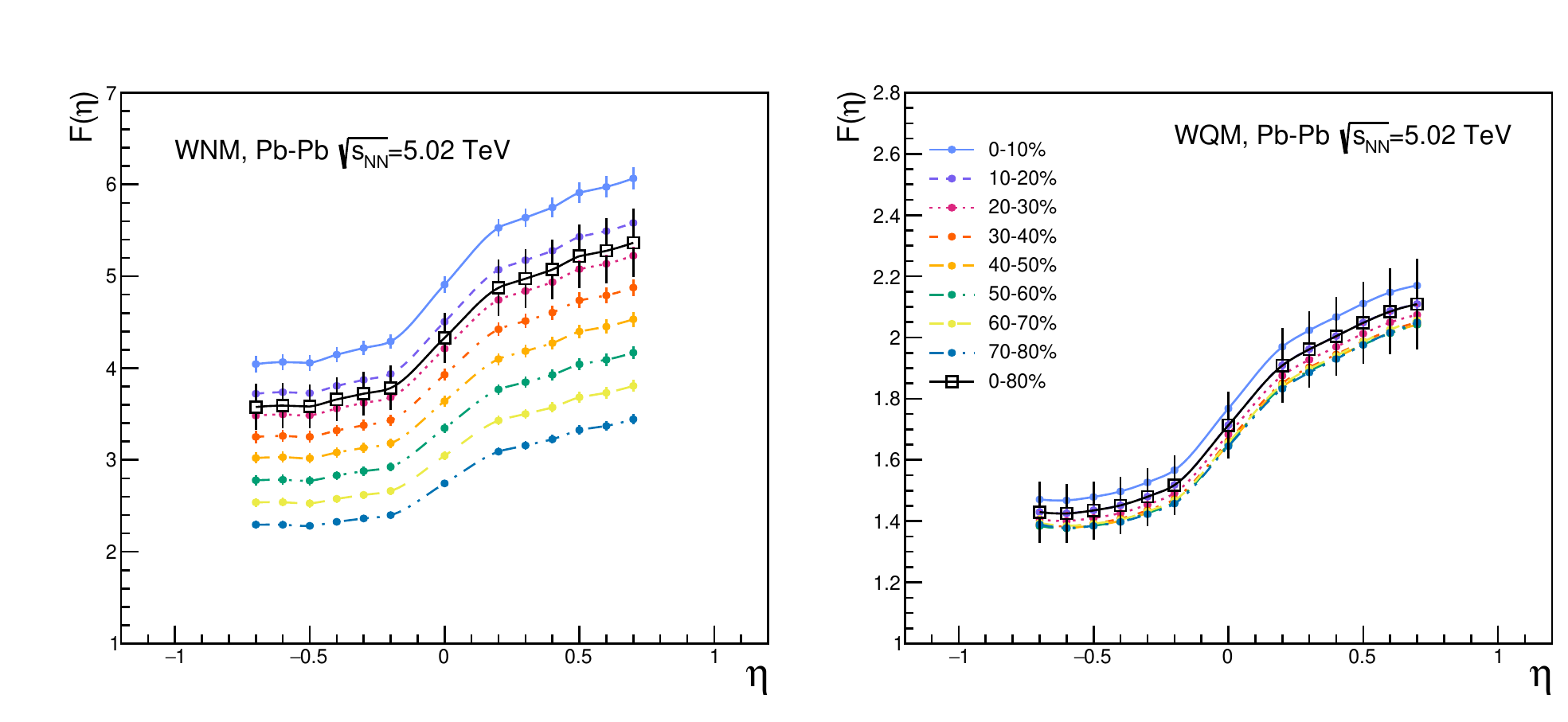}
\caption{\label{fig:fragmentation_functionPbPb} The  wounded nucleon (left panel) and quark (right panel) fragmentation functions extracted from symmetric Pb--Pb collisions at $\sqrt{s_{\rm{NN}}}$=5.02 TeV as a function of $\eta$ for different centrality classes. }
\end{figure*}

Figure~\ref{fig:fragmentation_functionPbPb} presents the fragmentation function of the wounded nucleon  and wounded quark determined in Pb--Pb collisions at $\sqrt{s_{\rm{NN}}}$=5.02 TeV.  Results are drawn as a function of $\eta$ for different centrality classes.
In figures, error bars mark total uncertainty of $F(\eta)$, to its values 
 contributed the uncertainty of the parameter $p$  and the uncertainty of $N(\pm\eta)$ determination  reported in Refs.~\cite{PbPbALICE:2016fbt}.

As seen from Fig.~\ref{fig:fragmentation_functionPbPb},  the shape of the wounded nucleon fragmentation function slightly changes with centrality class while its values increase with going from peripheral to central collisions. Whereas the fragmentation function of the wounded quark 
is more universal in nature;  namely, the shape and values are similar for all analyzed centrality classes. Qualitatively these findings coincide with the behavior of wounded nucleon and wounded quark fragmentation functions reported in the analysis of asymmetric d+Au  collisions at $\sqrt{s_{\rm{NN}}}$=200 GeV Ref.~\cite{Barej:2017kcw}. On the other hand, the fragmentation function determined in this work has a sharper trend with pseudorapidity than what is suggested by paper Ref.~\cite{Barej:2017kcw}. Further research is needed 
to understand this difference. 

\begin{figure}[h!]
\hspace*{-1.2cm} 
\includegraphics[trim=0 40 0 0,clip,width=0.59\textwidth]{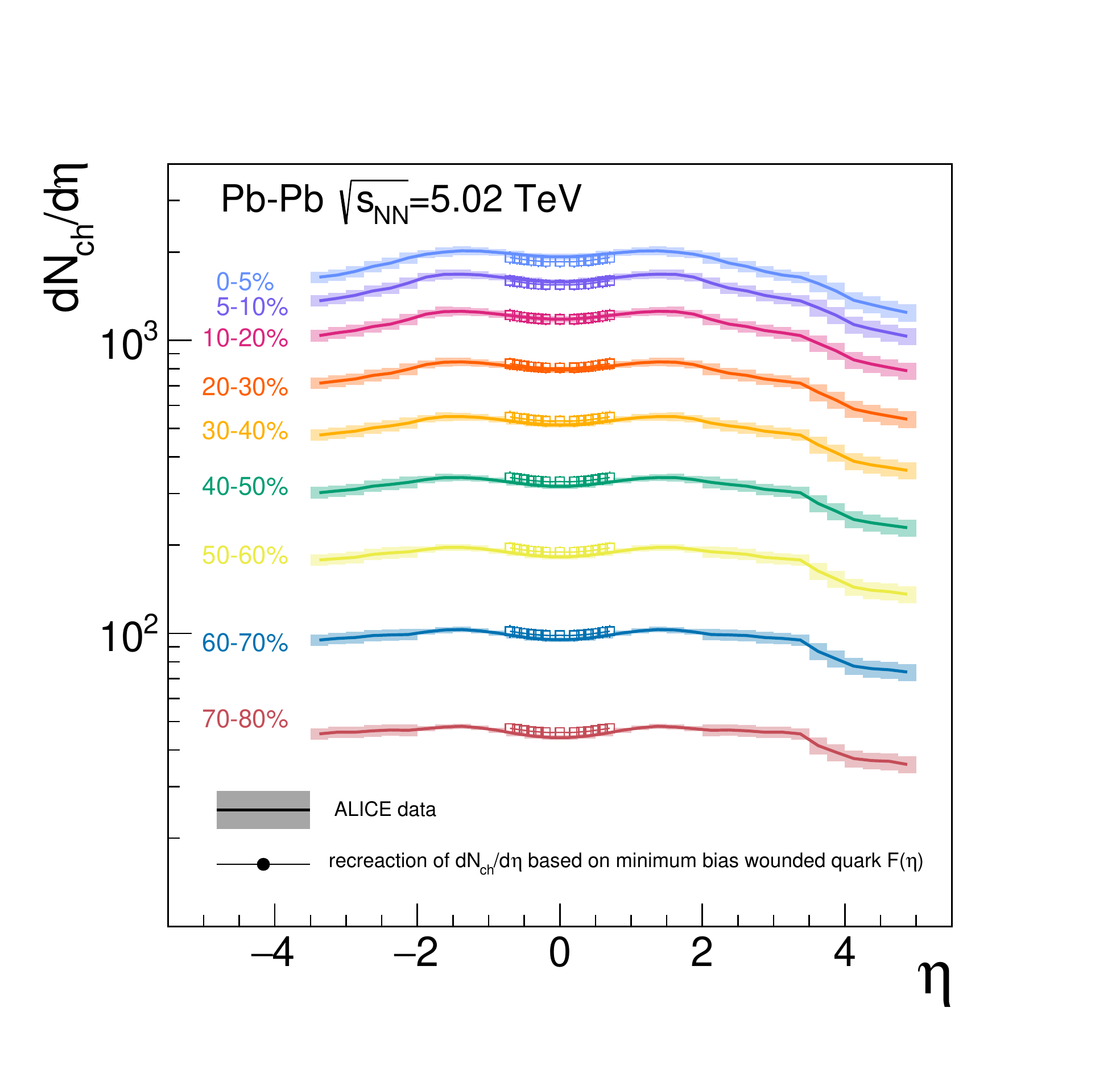}
\caption{\label{fig:dNdeta_distributionPbPb} Reconstruction of charged particle multiplicity distributions $N (\eta) \equiv  dN_{ch}/d\eta$
as functions of $\eta$ for Pb--Pb  collisions at $\sqrt{s_{\rm{NN}}}$=5.02 TeV using the extracted, based on the $\Sigma$ measurement, minimum bias  fragmentation function $F(\eta)$.
The data from ALICE is labeled with lines, while points represent this paper's calculation results. The corresponding uncertainties for the ALICE data and model calculations are denoted by shaded areas and bars, respectively. }
\end{figure}
As a cross-check exercise, the extracted minimum bias  fragmentation function of wounded quarks was used to determine charged particle density according to the Eq.~\ref{eq:particle_density} in  Pb--Pb collisions.  The results obtained on $dN_{ch}/d\eta$ were compared with the values measured in the ALICE experiment and are shown in Fig.~\ref{fig:dNdeta_distributionPbPb}. The figure shows good agreement between two sets of data points. 
\section{Conclusions}
This study set out to determine the properties of $\Sigma$ quantity at LHC energies in terms of two specific classes of independent source frameworks:  the wounded nucleon and the wounded quark model. The main findings to emerge from this study are listed below.
\begin{enumerate}
\item[1.] Incorporating two-component scenarios of forward- and backward-moving constituents makes the $\Sigma$ directly dependent on the number of wounded constituents, and thus collapses  the strongly intensive properties of this observable in the wounded nucleon and  wounded quark models.   
\item[2.] Even though in the WNM and WQM, $\Sigma$ is no longer a strongly intensive quantity, it retains some of its properties in symmetric collisions. In particular, it is free from volume fluctuations and does not depend on the method of centrality selection for $\leq10\%$ centrality bin width, which is consistent with the ALICE observation. This  emerges from the fact that, despite the width of the centrality interval and the method of centrality selection, $\Sigma$ always gives a covariance value between forward and backward wounded constituents at their fixed average number in a studied centrality sample.

\item[3.] The two considered wounded constituent models are in good agreement with the ALICE data. The WNM and WQM models depict the $\Sigma$ observable behavior that quite accurately matches the experimental results. These models outperform more complex ones such as HIJING, AMPT, or EPOS, which struggle to describe this observable properly, Refs.~\cite{Sputowska:2019yvr, Sputowska:2022CPOD, Sputowska:2022gai}.
\item[4.] One of the more surprising results from this analysis is the sensitivity of the $\Sigma$ quantity to the probability of particle emission in a given pseudorapidity interval by a wounded source. This relation allows the direct determination of the fragmentation function of a wounded nucleon or quark in a symmetric nucleus-nucleus collision, which has not been possible so far. 
\item[4.]  In this work,  the single particle wounded quark and nucleon fragmentation functions were the first-ever extracted for symmetric Pb--Pb collisions at LHC energies. 
\item[5.]The fragmentation function of the wounded quark has a more universal character and allows an accurate description of the charge-particle density distributions measured by ALICE.
\end{enumerate}

\begin{acknowledgments}
The author would like to thank Adam Bzdak for suggesting this investigation and useful discussions,  Andrzej Rybicki for valuable remarks and critical reading of the manuscript, and Wojciech Broniowski for introducing her to the concept of partial correlations. This work was supported by the National Science Center, Poland (grant No.2021/43/D/ST2/02195).
\end{acknowledgments}

\appendix

\section{Method of 
derivation of analytical formula on $\Sigma$} \label{app:generating}
In this Appendix, the calculation procedure of  the $\Sigma$ variable analytical formula in terms of  WNM and WQM is presented. 
The given procedure is based on the methodology proposed in the papers Refs.~\cite{Bzdak:2009dr, Bzdak:2009xq}, which use the generating function approach.

As derived in Refs.~\cite{Bzdak:2009dr, Bzdak:2009xq}  in the framework of the wounded constituent model, for forward  and backward pseudorapidity intervals symmetrically located around $\eta=0$, the generating function takes the form given by Eq.~\ref{eq:GF2}.
\begin{equation}
\begin{split}
        H(z_B,z_F) =&\sum_{w_F, w_B} W(w_F, w_B)\{1+ f\left(z_B,z_F\right)\}^{-\frac{k\times w_B}{2}}\times \\ 
    &\times\{1+g\left(z_B,z_F\right))\}^{-\frac{k\times w_F}{2}}\label{eq:GF2},\\
\end{split}
\end{equation} 
\begin{equation*}
\begin{split}
    \text{where:}\\
    &f\left(z_B,z_F\right) = \frac{\overline{n}}{k}\left[p(1-z_B)+q(1-z_F)\right]\\
   & g\left(z_B,z_F\right) = \frac{\overline{n}}{k}\left[q(1-z_B)+p(1-z_F)\right]\\
\end{split}
\end{equation*} 
In this formula, $W(w_F, w_B)$ represents the probability distribution of the numbers of wounded nucleons in the backward and forward-moving nucleus.
The quantities $p$ and $q=1-p$ denote the probabilities, where $p$ is defined by Eq.~\ref{eq:probability}.

By applying appropriate differentiation to the generating function defined in Eq.~\ref{eq:GF2}, one can determine the moments of the forward-backward multiplicity distribution:
\begin{eqnarray}\label{var_diff}
     \langle N_{F(B)}\rangle&=\eval{\pdv[1]{H(z_B,z_F)}{z_{F(B)}}}_{z_{F}=z_{B}=1}\\
     \langle N_{F(B)}^{2}\rangle&=\eval{\pdv[2]{H(z_B,z_F)}{z_{F(B)}}}_{z_{F}=z_{B}=1}\\
     \langle N_{F}N_{B}\rangle&=\eval{\pdv{H(z_B,z_F)}{z_{F}}{z_{B}}}_{z_{F}=z_{B}=1}.\\
\end{eqnarray}
Those moments are used to derive quantities such as $\omega_{F(B)}=\frac{Var(N_{F(B))}}{\langle N_{F(B)}\rangle}$, or $ \text{Cov($N_{B},N_{F}$)}=\langle N_{F}N_{B}\rangle- \langle N_{B}\rangle \langle N_{F}\rangle$ needed to construct $\Sigma$ according to Eq.~\ref{sigma_def}. 

\section{Derivation of the relation between $\Sigma$ quantity and partial covariance for fully symmetric collision}\label{app:cov}

To prove that for a fully symmetric collision, for which $\langle w_{B}\rangle =\langle w_{F} \rangle$ and $\text{Var} (w_{B}) =\text{Var}(w_{F})$,  the equation:
 \begin{equation}
\label{def}
  \boxed{ \text{Var}(w_{F}-w_{B}) =-4\text{Cov}(w_{F},w_{B} \bullet w)}
 \end{equation}
is satisfied,  we will use the following basic properties of variance and covariance:
\begin{equation}
Var(X-Y)=Var(X)+Var(Y)-2Cov(X,Y), \label{va}
\end{equation}
\begin{equation}
Var(X+Y)=Var(X)+Var(Y)+2Cov(X,Y), \label{vb}
\end{equation}
\begin{equation}
Cov(X,Y+Z)=Cov(X,Y)+Cov(X,Z). \label{vc}
\end{equation}
For $\langle w_{B}\rangle =\langle w_{F} \rangle$ we have:
\begin{eqnarray}\label{var_diff0}
   \boxed{ \langle (w_{F} -w_{B})^2 \rangle= \text{Var}(w_{F}-w_{B}),}
\end{eqnarray}
and by adding two expressions Eq.~\ref{va} and Eq.~\ref{vb}, the variance of the difference can be represented as:
\begin{eqnarray}\label{var_diff}
     \boxed{\text{Var}(w_{F}-w_{B})=\text{Var}(w_{F}+w_{B})-4\text{Cov}(w_{F},w_{B}).}\nonumber\\
\end{eqnarray}
Then based on properties of variance and covariance expressed in Eq.~\ref{vb} and Eq.~\ref{vc}, one can obtain:
    \begin{eqnarray}
    &\text{Var}(w_{F}+w_{B}) =\text{Var}(w_{F})+\text{Var}(w_{F})+2\text{Cov}(w_{F},w_{B})=\nonumber\\ 
    &= \underbrace{\text{Var}(w_{F})+\text{Cov}(w_{F},w_{B})}_{\text{Cov}(w_{F},w)}+\underbrace{\text{Var}(w_{B})+\text{Cov}(w_{F},w_{B})}_{\text{Cov}(w_{B},w)} =\nonumber\\
    &=\text{Cov}(w_{F},w)+\text{Cov}(w_{B},w) \label{var_diff}
    \end{eqnarray}
    where $w=w_{B}+w_{F}$.
\\For symmetric case $\text{Cov}(w_{F},w)=\text{Cov}(w_{B},w)$, so the above formula can be simplified to the form:
 \begin{equation}\label{var_diff_sym}
    \boxed{\text{Var}(w_{F}+w_{B}) =2\text{Cov}(w_{F},w).}
 \end{equation}
On the other hand, from the definition of partial covariance (Eq.~\ref{partial_cov}) we find: 
\begin{equation}
\label{p_cov}
 \text{Cov}(w_{F},w_{B} \bullet w) = \text{Cov}(w_{F},w_{B})-\frac{\text{Cov}(w_{F},w) \text{Cov}(w_{B},w) }{\text{Var}(w)},
 \end{equation}
 which for the symmetric collision takes the form:
\begin{equation}
\label{p_cov_sym}
 \boxed{\text{Cov}(w_{F},w_{B} \bullet w) = \text{Cov}(w_{F},w_{B})-\frac{\text{Cov}(w_{F},w)^2 }{\text{Var}(w)}.}
 \end{equation}

Combining Eq.~\ref{p_cov_sym} and Eq.~\ref{var_diff_sym} we get:
\begin{eqnarray}
\label{cov_sym}
 &\text{Cov}(w_{F},w_{B} \bullet w) = \text{Cov}(w_{F},w_{B})-\frac{\text{Var}(\overbrace{w_{F}+w_{B}}^{w})^2 }{4\text{Var}(w)}=\nonumber\\
 &=\text{Cov}(w_{F},w_{B})-\frac{\text{Var}(w)}{4}
 \end{eqnarray}
 And finally, transforming the Eq.~\ref{cov_sym} and taking into account Eq.~\ref{var_diff} we find the relation given by Eq.~\ref{FLUC_COV}:
 \begin{eqnarray}
 \label{cov_final_sym}
 \boxed{-4\text{Cov}(w_{F},w_{B} \bullet w) = \underbrace{-4\text{Cov}(w_{F},w_{B})+\text{Var}(w)}_{\text{Var}(w_{F}-w_{B})}}\nonumber\\
 \end{eqnarray}

\nocite{*}

\bibliography{apssamp}

\end{document}